\documentclass[11pt]{article}
\usepackage{amsmath,amssymb}

\makeatletter
\@addtoreset{equation}{section}
\makeatother

\def\nn{\nonumber} \def\bd{\begin{document}} \def\ed{\end{document}}
\def\ds{\documentstyle}
\let\bm=\bibitem
\newcommand{\be}{\begin{equation}}
\newcommand{\ee}{\end{equation}}
\newcommand{\bea}{\setlength\arraycolsep{2pt} \begin{eqnarray}}
\newcommand{\eea}{\end{eqnarray}}
\newcommand{\hoch}[1]{$\, ^{#1}$}
\newcommand{\auth}{
Jun-Jin Peng\hoch{\dagger}, Shuang-Qing Wu
}

\begin{document}

\vspace{25pt}
\begin{center}
{\large {\bf Extremal Kerr black hole/CFT correspondence in the
five-dimensional G\"{o}del universe}}

\vspace{15pt}
\auth

\vspace{20pt}
{\it College of Physical Science and Technology, Central China Normal
University,\\ Wuhan, Hubei 430079, People's Republic of China}

\vspace{20pt}
E-Mail: pengjjph@163.com\hoch{\dagger}
\vspace{80pt}

\textbf{Abstract}  
\end{center}
We extend the method of Kerr/CFT correspondence recently proposed in arXiv:0809.4266
[hep-th] to the extremal (charged) Kerr black hole embedded in the five-dimensional
G\"{o}del universe. With the aid of the central charges in the Virasoro algebra and
the Frolov-Thorne temperatures, together with the use of the Cardy formula, we have
obtained the microscopic entropies that precisely agree with the ones macroscopically
calculated by Bekenstein-Hawking area law.

\vspace{15pt}

\thispagestyle{empty}

\pagebreak
\setcounter{page}{1}

%%%%%%%%%%%%%%%%%%%%%%%%%%%%%%%%%%%%%%%%

\newpage
\section{Introduction}

 During the past decades, a lot of efforts have been devoted to studying the origin
of Bekenstein-Hawking entropy for the black holes. Great progress has been made in
their statistical interpretation thanks to Strominger and Vafa's remarkable work
\cite{strova} on the investigation of the microscopic origin of the five-dimensional,
supersymmetric (extremal) black hole entropy by using the holographic duality in
the context of string theory. When the near-horizon limit has been taken, their
work can be viewed as a typical example of the AdS/CFT correspondence
\cite{jmaldacena,gubserkp,ewitten}, which shows that there exists a duality between
the higher dimensional gravity and the CFT living on the boundary in less dimensions,
providing a powerful tool to study the microscopic statistical mechanics of the
black holes. By contrast, without using any supersymmetry, Strominger \cite{strominger}
had successfully evaluated the Bekenstein-Hawking entropy of the three-dimensional
BTZ black hole by counting the number of the microstates in the two-dimensional CFT
induced on the boundary of spatial infinity \cite{JDbrown}. Besides, when the boundary
is restricted to the horizon, the statistical entropy of the general
black hole was derived via studying the relationship between black hole thermodynamics
and the  two dimensional near-horizon CFT \cite{carlip,solod,muinpark,kkpark}.

Quite recently, Guica, Hartman, Song and Strominger \cite{GHSStro} put forward a new
method called as Kerr/CFT correspondence to derive the microscopic entropy of the
four-dimensional extremal Kerr black hole by identifying the quantum states in its
near-horizon region with the two-dimensional chiral CFT on the spatially infinite
boundary. The main ideas of their method go as follows: On the basis of the near-horizon
geometry found in \cite{Wang,Bardeen}, one can construct diffeomorphisms that preserve
a properly chosen boundary condition at the infinity. These diffeomorphisms generate
one copy of the Virasoro algebra and contribute to the conserved charges. By computing
the Dirac brackets of these charges, the central charge relative to the angular momentum
of the extremal black hole can be obtained. Making use of the Frolov and Thorne temperature
\cite{Frothorne}, the microscopic entropy in the dual CFT can be reproduced via the Cardy
formula. Following this work, the microscopic entropies of the three-dimensional black hole
and Kerr-AdS black holes in diverse dimensions were derived \cite{hotta,lumeip}. In Refs.
\cite{Azeyanagi,HKNStro}, the Kerr/CFT correspondence was applied to the black holes with
$U(1)$ gauge symmetry. Further extensions \cite{chowlv,Nakayama,Isonotai,Ogawa,cmchen}
has been made in supergravity theory and string theory.

In this paper, we shall apply the Kerr/CFT correspondence to the extremal (charged) Kerr
black hole embedded in the five-dimensional G\"{o}del universe \cite{Gimon,sqwu}, which is
dubbed as a (charged) Kerr-G\"{o}del black hole for shortness. These black hole metrics are
exact solutions in the five-dimensional minimal supergravity. They possess some peculiar
features such as the presence of closed time-like curves, and the absence of globally
spatial-like Cauchy surface. Our paper is organized as follows. In Section \ref{two},
we simply review the Kerr-G\"{o}del black hole and obtain its near-horizon metric under
the extremal condition. Based upon the near-horizon metric, we then calculate the central
charge and microscopic entropy of the extremal Kerr-G\"{o}del black hole in the chiral dual
CFT. In Section \ref{three}, we extend the analysis to the extremal charged Kerr-G\"{o}del
black hole. Finally, in Section \ref{four}, we end up with our conclusions.

%%%%%%%%%%%%%%%%%%%%%%%%%%%%%%%%%%%%%%%%%%%%%%%%%%%%%%%%%%%%%%%%%%%%%%%%
\section{Extremal Kerr-G\"{o}del black hole and the dual CFT}\label{two}
%%%%%%%%%%%%%%%%%%%%%%%%%%%%%%%%%%%%%%%%%%%%%%%%%%%%%%%%%%%%%%%%%%%%%%%%

In this section, we will generalize the method developed in \cite{GHSStro,HKNStro} to explore
the duality between the extremal Kerr-G\"{o}del black hole \cite{Gimon} and the chiral CFT by
showing the equality of the microscopic CFT entropy and the Bekenstein-Hawking entropy. We
first give a brief review of the Kerr-G\"{o}del black hole solution \cite{Gimon} and then
study its near-horizon geometry. Let's start with the metric
\bea
ds^2 &=& -\big(1 -\frac{2\mu}{\hat{r}^2}\big)d\hat{t}^2
 +\frac{d\hat{r}^2}{\Delta_{\hat{r}}} -4\big(j\hat{r}^2 +\frac{\mu a}{\hat{r}^2}\big)
 (\cos^2\theta d\hat{\phi} +\sin^2\theta d\hat{\psi})d\hat{t} \nn \\
 && -4\hat{r}^2\big(j^2\hat{r}^2 +2j^2\mu -\frac{\mu a^2}{2\hat{r}^4}\big)(\cos^2\theta
 d\hat{\phi} +\sin^2\theta d\hat{\psi})^2 \nn \\
 && +\hat{r}^2(d\theta^2 +\cos^2\theta d\hat{\phi}^2 +\sin^2\theta d\hat{\psi}^2) \, ,
\label{kerrGmetric}
\eea
and the gauge potential which takes the form
\be
A = \sqrt{3}j\hat{r}^2(\cos^2\theta d\hat{\phi} +\sin^2\theta d\hat{\psi}) \, ,
\ee
where
\be
\Delta_{\hat{r}} = 1 -\frac{2\mu}{\hat{r}^2} +\frac{16j^2\mu^2}{\hat{r}^2}
 +\frac{8j\mu a}{\hat{r}^2} +\frac{2\mu a^2}{\hat{r}^4} \, .
\ee

In the above, the parameters $\mu$ and $a$ are related to the mass and the angular momenta,
respectively, while $j$ defines the scale of the G\"{o}del background and is responsible for
the rotation of the universe. Without loss of generality, we assume $\mu$, $a$ and $j$ are
all positive. The metric (\ref{kerrGmetric}) describes the rotating black hole with two equal
angular velocities in the five-dimensional G\"{o}del universe. The angular velocities
and the electro-static potential on the horizon are given by
\be
\Omega^H_{\hat{\phi}} = \Omega^H_{\hat{\psi}}
 = 2(j\hat{r}_H^4 +\mu a)/\eta \, , \qquad
\Phi_H = -2\sqrt{3}j\hat{r}_H^2(j\hat{r}_H^4 +\mu a)/\eta \, .
\ee
where the event horizon $\hat{r}_H$ and constant $\eta$ read
\bea
\hat{r}_H^2 &=& \mu -4\mu ja -8j^2\mu^2 +\mu\sqrt{(1 -8\mu j^2)(1 -8\mu j^2
 -8ja -2\mu^{-1}a^2)} \, , \nn \\
\eta &=& \hat{r}_H^4 +2\mu a^2 -4j^2\hat{r}_H^6 -8\mu j^2\hat{r}_H^4 \, . \nn
\eea
The temperature and the entropy are
\be
S = 2\pi^2\hat{r}_H\sqrt{\eta} \, , \qquad
T_H = \frac{\hat{r}_H^2 -\mu +4j\mu a +8j^2\mu^2}{\pi\hat{r}_H \sqrt{\eta}} \, .
\ee
In \cite{Barnichprl}, the conserved quantities such as the mass, the angular momenta
and the electrical charge have been computed as
\bea
M &=& \frac{3}{4}\pi \mu -8\pi j^2\mu^2 -\pi j\mu a \, , \nn \\
J_{\hat{\phi}} &=& J_{\hat{\psi}} = \frac{1}{2}\pi\mu a
 -\pi j\mu a^2 -4\pi aj^2\mu^2 \, , \\
\mathcal{Q} &=& 2\sqrt{3}\pi j\mu a \, . \nn
\eea
Treating the G\"{o}del parameter $j$ as a fixed constant, we find that the variation
of the mass, the angular momenta and the electrical charge satisfy the differential
form of the first law
\be
dM = T_H dS +\Omega^H_{\hat{\phi}}dJ_{\hat{\phi}}
 +\Omega^H_{\hat{\psi}}dJ_{\hat{\psi}} +\Phi_H d\mathcal{Q} \, .
\ee
However, if we take $j$ as a thermodynamical variable, a conjugate generalized force
should be introduced \cite{sqwu} to fulfill the first law of the black hole thermodynamics.

Now we turn our attention to the analysis of the near-horizon geometry of the extremal
Kerr-G\"{o}del black hole. The extremity condition is
\be
j = \frac{(\mu -r_0^2)\sqrt{2}}{4\mu^{3/2}} \, , \qquad
a = \frac{r_0^2}{\sqrt{2\mu}} \, ,
\label{extrecondi}
\ee
where $r_0$ is the horizon radius of the extremal black hole, which makes the temperature
$T_H$ vanish. Under the extremal condition (\ref{extrecondi}), to obtain the near-horizon
geometry of the Kerr-G\"{o}del black hole, we perform the coordinate transformation as
follows
\bea
\hat{r} &=& r_0 +r_0\lambda r \, , \qquad\qquad~~
\hat{t} = \frac{r_0(\mu +r_0^2)\sqrt{2(2\mu -r_0^2)}}{8\mu^{3/2}\lambda}t \, , \nn \\
\hat{\phi} &=& \phi +\frac{\sqrt{2\mu -r_0^2}}{4r_0\lambda}t \, , \qquad
\hat{\psi} = \psi +\frac{\sqrt{2\mu -r_0^2}}{4r_0\lambda}t \, ,
\eea
and then take the scaling parameter $\lambda$ to zero, thus sending the metric
(\ref{kerrGmetric})
to the form
\bea
&&\hspace*{-0.7cm}
ds^2 = \frac{1}{4}r_0^2\big(-r^2dt^2 +\frac{dr^2}{r^2} +4d\theta^2\big)
 -\frac{r_0^4(3\mu^2 -r_0^4)}{2\mu^3}
 \cos^2\theta \sin^2\theta(d\phi -d\psi)^2 \nn \\
&&\quad +\frac{r_0^2(2\mu -r_0^2)(\mu +r_0^2)^2}{2\mu^3} \big[\cos^2\theta
 (d\phi +\alpha rdt)^2 +\sin^2\theta (d\psi+\alpha rdt)^2\big] \, ,
\label{exmetric}
\eea
where
\be
\alpha = \frac{r_0(3\mu -r_0^2)}{2(\mu +r_0^2)\sqrt{2\mu -r_0^2}} \, .
\ee
The near-horizon metric (\ref{exmetric}) depicts a 3-sphere bundle over the $AdS_2$ space.
It only partially covers the near-horizon geometry of the extremal Kerr-G\"{o}del black hole
(\ref{kerrGmetric}). One can perform global coordinate transformation to the coordinates
$(t, r)$ in order to make the metric (\ref{exmetric}) overlay the whole space in a single
patch \cite{Bardeen,chowlv}.

By virtue of the conformal structure of the near-horizon metric (\ref{exmetric}), it is possible
for us to compute the central charges in the chiral CFT. Since there exist two rotations
corresponding to $\phi$ and $\psi$, respectively, when the-near horizon metric (\ref{exmetric})
is assumed to have a certain suitable boundary, it can be shown as did in \cite{GHSStro} that
this metric can possess two commuting diffeomorphisms
\bea
\zeta_{n}^{(1)} &=& -e^{-\mathrm{i}n\phi}\partial_\phi
 -\mathrm{i}nre^{-\mathrm{i}n\phi}\partial_r \, , \nn \\
\zeta_{n}^{(2)} &=& -e^{-\mathrm{i}n\psi}\partial_\psi
 -\mathrm{i}nre^{-\mathrm{i}n\psi}\partial_r \, ,
\qquad (n = 0, \pm 1, \pm 2, ...)
\label{comdeff}
\eea
which preserve the chosen boundary and generate two copies of commuting Virasoro algebra
\be
\mathrm{i}\big[\zeta_{m}^{(i)}, \zeta_{n}^{(j)}\big]
 = (m -n)\delta_{ij}\zeta_{m+n}^{(i)} \qquad (i, j = 1, 2) \, .
\ee
Each diffeomorphism $\zeta_{m}^{(i)}$ is associated to a conserved charge defined by
\cite{BarBran,BarCjmp,Compere,Barcqg}
\be
Q_{\zeta_{n}^{(i)}} =\frac{1}{8\pi} \int_{\partial\Sigma}k_{\zeta_{n}^{(i)}}[h, g] \, ,
\ee
where $\partial\Sigma$ is a spatial slice that extends to the infinity and the 3-form
$k_{\zeta_{n}^{(i)}}[h, g]$ is written as
\bea
k_{\zeta}[h, g] &=& -\frac{1}{12}\epsilon_{\alpha\beta\gamma\rho\sigma}
 \big[\zeta^\rho\nabla^\sigma h -\zeta^\rho\nabla_\nu h^{\sigma\nu}
 +\zeta_\nu\nabla^\rho h^{\sigma\nu} +\frac{1}{2}h\nabla^\rho\zeta^\sigma \nn \\
&& -h^{\rho\nu}\nabla_\nu\zeta^\sigma +\frac{1}{2}h^{\rho\nu}(\nabla^\sigma\zeta_\nu
 +\nabla_\nu\zeta^\sigma)\big]dx^\alpha\wedge dx^\beta\wedge dx^\gamma \, .
\eea
In the above equation, $\zeta=\zeta_{n}^{(i)}$, and $h_{\rho\sigma}$ denotes the deviation
from the background metric (\ref{exmetric}). The Dirac brackets of the conserved charges
corresponding to the diffeomorphisms $\zeta_{m}^{(i)}$ and $\zeta_{n}^{(i)}$ yield the
central term in the Virasoro algebra
\be
\frac{1}{8\pi}\int_{\partial\Sigma}k_{\zeta_{m}^{(i)}}[\mathcal{L}_{\zeta_{n}^{(i)}}g, g]
= -\frac{\mathrm{i}}{12}c_i(m^3 +\beta m)\delta_{m+n, 0} \, ,
\label{centerm}
\ee
where
\be
\mathcal{L}_{\zeta_{n}^{(i)}}g_{\rho\sigma} = \zeta_{n}^{(i)\nu}
 \partial_\nu g_{\rho\sigma} +g_{\nu\sigma}\partial_\rho
 \zeta_{n}^{(i)\nu} +g_{\nu\rho}\partial_\sigma\zeta_{n}^{(i)\nu}
\ee
is the Lie derivative of the background metric (\ref{exmetric}) relative to the vector field
$\zeta_{n}^{(i)}$, $c_i$ denote the central charges in the Virasoro algebra, while the constant
$\beta$ is trivial since it can be absorbed by a shift in $Q_{\zeta_{0}^{(i)}}$. In terms of
the background metric (\ref{exmetric}), the quantities associated with the central charges
$c_i$ are
\bea
\frac{1}{8\pi}\int_{\partial\Sigma}k_{\zeta_{m}^{(i)}}[\mathcal{L}_{\zeta_{n}^{(i)}}g, g]
&=& -\frac{\mathrm{i}\pi r_0^3(\mu +r_0^2)\sqrt{2(2\mu -r_0^2)}}{8\mu^{9/2}}
 \alpha[\mu^3m^3 \nn \\
&& +(2\mu -r_0^2)(\mu +r_0^2)^2m]\delta_{m+n, 0} \, .
\label{crecent}
\eea
Comparing Eq. (\ref{crecent}) with (\ref{centerm}), we get the central charges in the chiral
Virasoro algebra
\be
c_1 = c_2 = \frac{3\pi r_0^3(\mu +r_0^2)\sqrt{2(2\mu -r_0^2)}}{2\mu^{3/2}}\alpha
\label{centralchar} \, .
\ee

With the central charges (\ref{centralchar}) in hand, we have to determine the Frolov-Thorne
temperatures to calculate the chiral CFT entropy via the Cardy formula. Let $T_1$ and $T_2$
represent the Frolov-Thorne temperatures associated with the azimuthal angles $\phi$ and $\psi$,
respectively. Using the definition of the Frolov-Thorne temperature in \cite{chowlv}, we
obtain
\be
T_1 = T_2 = -\lim_{\hat{r}_H\rightarrow r_0} \frac{T_H}{\Omega^H_{\hat{\psi}}
 -\Omega^0_{\hat{\psi}}} = \frac{1}{2\pi \alpha} \, ,
\ee
where
\be
\Omega^0_{\hat{\psi}} = \Omega^H_{\hat{\psi}}(T_H = 0)
 = \frac{\sqrt{2\mu^3}}{r_0^2(\mu +r_0^2)}
\ee
is the angular velocity of the extremal Kerr-G\"{o}del black hole. Finally,
with the aid of the Cardy formula for the microscopic entropy in the chiral
CFT, we get
\be
S_1 = S_2 = \frac{\pi^2}{3}c_2T_2
 = \frac{\pi^2r_0^3(\mu +r_0^2)\sqrt{2(2\mu -r_0^2)}}{4\mu^{3/2}} \, ,
\ee
which precisely agree with the macroscopic Bekenstein-Hawking entropy of the extremal
Kerr-G\"{o}del back hole
\be
S(T_H = 0) = \frac{\pi^2r_0^3(\mu +r_0^2)\sqrt{2(2\mu -r_0^2)}}{4\mu^{3/2}} \, .
\ee

%%%%%%%%%%%%%%%%%%%%%%%%%%%%%%%%%%%%%%%%%%%%%%%%%%%%%%%%%%%%%%%%%%%%%%%%%%%%%
\section{Extremal charged Kerr-G\"{o}del black hole and CFT duality}\label{three}
%%%%%%%%%%%%%%%%%%%%%%%%%%%%%%%%%%%%%%%%%%%%%%%%%%%%%%%%%%%%%%%%%%%%%%%%%%%%%

In this section, we will extend the above analysis to the extremal charged Kerr black hole
in the G\"{o}del universe, which is an exact solution \cite{sqwu} in the five-dimensional
Einstein-Maxwell-Chern-Simons supergravity theory. Parameterized by four constants $(\mu,
a, q, j)$, which correspond to the mass, the angular momenta, the electric charge and the
scale of the G\"{o}del background, respectively, the metric has the form
\bea
ds^2 &=& -\frac{\hat{r}^2V(\hat{r})}{4B(\hat{r})}d\hat{t}^2 +\frac{d\hat{r}^2}{V(\hat{r})}
 +\hat{r}^2[d\theta^2 +\cos^2\theta \sin^2\theta(d\hat{\phi} -d\hat{\psi})^2] \nn \\
&& +4B(\hat{r})\Big(\cos^2\theta d\hat{\phi} +\sin^2\theta d\hat{\psi}
 -\frac{1}{2}\frac{G(\hat{r})}{B(\hat{r})}d\hat{t}\Big)^2 \, ,
\label{charmetric}
\eea
and the gauge potential is given as
\be
A = \frac{\sqrt{3}q}{2\hat{r}^2}d\hat{t} +\sqrt{3}(j\hat{r}^2 +2jq -\frac{aq}{2\hat{r}^2})
(\cos^2\theta d\hat{\phi} +\sin^2\theta d\hat{\psi}) \, ,
\ee
where
\bea
G(\hat{r}) &=& j\hat{r}^2 +3jq +\frac{(2\mu -q)a}{2\hat{r}^2}
 -\frac{q^2a}{2\hat{r}^4} \, ,\nn \\
B(\hat{r}) &=& -j^2\hat{r}^2(\hat{r}^2 +2\mu +6q) +3jqa +\frac{(\mu-q)a^2}{2\hat{r}^2}
 -\frac{q^2a^2}{4\hat{r}^4} +\frac{\hat{r}^2}{4} \, , \nn \\
V(\hat{r}) &=& 1 -\frac{2\mu}{\hat{r}^2} +\frac{8j(\mu +q)[a +2j(\mu +2q)]}{\hat{r}^2} \nn \\
&& +\frac{2(\mu -q)a^2 +q^2[1 -16ja -8j^2(\mu+3q)]}{\hat{r}^4} \, .
\eea
The line element (\ref{charmetric}) is the charged generalization of the metric
(\ref{kerrGmetric}). On the choice of proper parameters, it can cover other solutions.
For example, when $j = 0$, it becomes the Kerr-Newman black hole solution with two equal
rotations in the five-dimensional minimal supergravity. If we assume $j = 0$ and $q = m$,
the metric (\ref{charmetric}) reduces to the BMPV black hole \cite{BMPV,Kallosh}, whose
microscopic entropy has been recently obtained \cite{Isonotai,Ogawa,cmchen} via the method of
Kerr/correspondence. Now we present some useful quantities related to the charged
Kerr-G\"{o}del black hole (\ref{charmetric}), such as the angular velocities, the
temperature and the Bekenstein-Hawking entropy. We have
\be
\Omega_{\hat{\phi}} = \Omega_{\hat{\psi}} = \frac{G(\hat{r}_+)}{2B(\hat{r}_+)} \, , \qquad
T_H = \frac{\hat{r}_+V'(\hat{r}_+)}{8\pi\sqrt{B(\hat{r}_+)}} \, , \qquad
S = \pi^2\hat{r}^2_+\sqrt{B(\hat{r}_+)} \, ,
\ee
in which and what follows, the prime $'$ denotes the derivative with respect to the coordinate
$\hat{r}$. The event horizon $\hat{r}_+$ is determined by the equation $V(\hat{r}_+) = 0$ and
reads
\bea
\hat{r}^2_+ &=& \mu -4j(\mu +q)a -8j^2(\mu +q)(\mu +2q) +\sqrt{\delta} \, , \nn \\
\delta &=& [\mu -q -8j^2(\mu +q)^2] \nn \\
&& \times[\mu +q -2a^2 -8j(\mu +2q)a -8j^2(\mu +2q)^2] \, .
\eea
Obviously, $\delta > 0$ guarantees that the event horizon is well defined. However, when
$\delta = 0$, the event horizon degenerates to the extremal case, in which the charged
Kerr-G\"{o}del black hole has zero temperature but finite entropy
\be
S(T_H = 0) = \pi^2r^2_0\sqrt{B(r_0)} \, ,
\label{charmaentr}
\ee
where $r_0$ is the event horizon of the extremal charged Kerr-G\"{o}del black hole.

As before, we are interested in studying the equivalence property between the Bekenstein-Hawking
entropy and the one of the extremal charged Kerr-G\"{o}del black hole in the chiral CFT. To do
so, we first perform the coordinate transformation
\bea
\hat{r} &=& r_0 +r_0\lambda r \, , \qquad\quad~~
\hat{t} = \frac{2\sqrt{B(r_0)}}{\omega r_0^2}~\frac{t}{\lambda} \, , \nn \\
\hat{\phi} &=& \phi +\frac{G(r_0)}{2B(r_0)}\hat{t} \, , \qquad
\hat{\psi} = \psi +\frac{G(r_0)}{2B(r_0)}\hat{t} \, ,
\eea
where $\omega=V''(r_0)/2$, and then take $\lambda\rightarrow 0$ to get
the near-horizon metric
\bea
ds^2 &=& \frac{1}{\omega}\Big(-r^2dt^2 +\frac{dr^2}{r^2}\Big)
 +r_0^2\big[d\theta^2 +\cos^2\theta \sin^2\theta(d\phi -d\psi)^2\big] \nn \\
&&+4B(r_0)(\cos^2\theta d\phi +\sin^2\theta d\psi +krdt)^2 \, ,
\label{exchametr}
\eea
in which the constant $k$ is defined by
\be
k = \frac{B'(r_0)G(r_0) -G'(r_0)B(r_0)}{\omega r_0B^{3/2}(r_0)} \, .
\ee
Next, for the background metric (\ref{exchametr}), we directly calculate the central charges
in the chiral Virasoro algebra following the procedure in the previous section. The central
terms associated with the diffeomorphisms $\zeta_{n}^{(1)}$ and $\zeta_{n}^{(2)}$ can be read
off as
\be
\frac{1}{8\pi}\int_{\partial\Sigma}k_{\zeta_{m}^{(i)}}[\mathcal{L}_{\zeta_{n}^{(i)}}g, g]
 = -\frac{\mathrm{i}}{2}\pi k r_0^2\sqrt{B(r_0)} [m^3 +2\omega B(r_0)m]\delta_{m+n, 0} \, .
\label{charcent}
\ee
By comparison with Eq. (\ref{centerm}), we obtain the central charges
\be
c_1 = c_2 = 6k\pi r_0^2\sqrt{B(r_0)} \, .
\label{charcen}
\ee
The Frolov-Thorne temperatures of the chiral CFT can be expressed in the forms
\be
T_1 = T_2 = -\frac{\partial_{r_+} T_H}{\partial_{r_+} \Omega_{\hat{\phi}}}
\Big|_{r_+ = r_0} = \frac{1}{2\pi k} \, .
\label{ftcharte}
\ee
Finally, substituting Eqs. (\ref{charcen}) and (\ref{ftcharte}) into the Cardy formula,
we can compute the microscopic entropies in the chiral CFT as
\be
S_1 = S_2 = \frac{\pi^2}{3}c_2T_2 =\pi^2r^2_0\sqrt{B(r_0)}  \, .
\label{chacftentr}
\ee
Clearly, $S_1$ and $S_2$ are consistent with the one of the macroscopically calculated
entropy (\ref{charmaentr}). Note that the derivation of Eq. (\ref{chacftentr}) heavily
relies on the non-vanishing angular velocities of the horizon. For example, if we choose
a special but nontrivial relation of the parameters, namely \cite{sqwu}
\be
(\mu -a)a^2 +4j(\mu -q)(\mu +2q)a -4j^2(3\mu +5q)q^2 = 0 \, ,
\ee
which leads to $V(r_+) = g(r_+) = 0$, the Kerr/CFT correspondence fails to derive the
microscopic entropies in Eq. (\ref{chacftentr}). Besides, when $j = 0$ and $m = q$,
carrying out the coordinate transformation
\be
\hat{\phi} \rightarrow \phi+\psi \, , \quad
\hat{\psi} \rightarrow \phi-\psi \, , \quad
\theta \rightarrow \theta /2 \, ,
\label{newthpp}
\ee
to the metric (\ref{charmetric}), we get the BMPV black hole  metric that takes the same
form as the one in \cite{Isonotai}. Our derivation reproduces the microscopic entropy
there.

%%%%%%%%%%%%%%%%%%%%%%%%%%%%%%
\section{Summary}\label{four}
%%%%%%%%%%%%%%%%%%%%%%%%%%%%%%

In this paper, we have derived the microscopic entropy of a five-dimensional extremal
(charged) Kerr-G\"{o}del black hole by utilizing the Kerr/CFT correspondence recently
suggested in \cite{GHSStro}. Making use of the near-horizon limit procedure \cite{Bardeen},
we got the expected near-horizon metrics of the extremal black holes. They possess the
topological structure of a 3-sphere bundle over $AdS_2$. Choosing proper boundaries due to
the near-horizon metrics, we have two diffeomorphisms $(\zeta_{n}^{(1)}, \zeta_{n}^{(2)})$
that preserve these boundaries. Both the diffeomorphisms generate two copies of commuting
Virasoro algebra but the Dirac brackets of their corresponding conserved charges induce
non zero central terms in the Virasoro algebra. From the central terms, we can obtain the
central charges. In favor of the Frolov-Thorne temperatures, the entropies of the extremal
(charged) Kerr-G\"{o}del black hole were calculated via the Cardy formula. The entropies
got in the chiral dual CFT take the same value as the ones of the Hawking-Bekenstein entropy.
Our work implies that the Kerr/CFT correspondence is valid in the background of the
five-dimensional G\"{o}del universe, and recovers the previous results
\cite{Isonotai,Ogawa,cmchen} in some special cases.

\newpage
\section*{Acknowledgments}

S.Q.-Wu was partially supported by the Natural Science Foundation of China under
Grant No. 10675051.

\end{document}